# Exploring Domain Wall Pinning in Ferroelectrics via Automated High-Throughput AFM

Kamyar Barakati[1, *], Yu Liu[1], Hiroshi Funakubo[2], and Sergei V. Kalinin[1,3, †]

[1] Department of Materials Science and Engineering, University of Tennessee, Knoxville, TN 37996

[2] Department of Materials Science and Engineering, Institute of Science Tokyo, Yokohama 226-8502, Japan

[3] Pacific Northwest National Laboratory, Richland, WA 99354

**Abstract:**

Domain-wall dynamics in ferroelectric materials are strongly position-dependent since each polar interface is locked into a unique local microstructure. This necessitates spatially resolved studies of the wall-pinning using scanning-probe microscopy techniques. The pinning centers and preexisting domain walls are usually sparse within image plane, precluding the use of dense hyperspectral imaging modes and requiring time-consuming human experimentation. Here, a large-area epitaxial $PbTiO_3$ film on cubic $KTaO_3$ were investigated to quantify the electric-field-driven dynamics of the polar–strain domain structures using ML-controlled automated Piezoresponse Force Microscopy. Analysis of $1500$ switching events reveals that domain-wall displacement depends not only on field parameters but also on the local ferroelectric–ferroelastic configuration. For example, twin boundaries in polydomains regions like $\boldsymbol{a}_1^-/\boldsymbol{c}^+ \parallel \boldsymbol{a}_2^-/\boldsymbol{c}^-$ stay pinned up to a certain level of bias magnitude and change only marginally as the bias increases from $20\,V$ to $30\,V$, whereas single-variant boundaries like $\boldsymbol{a}_2^+/\boldsymbol{c}^+ \parallel \boldsymbol{a}_2^-/\boldsymbol{c}^-$ stack are already activated at $20\,V$. These statistics on the possible ferroelectric and ferroelastic wall orientations, together with the automated, high-throughput AFM workflow, can be distilled into a predictive map that links domain configurations to pulse parameters. This microstructure-specific rule set forms the foundation for designing ferroelectric memories.

* kbarakat@vols.utk.edu

† sergei2@utk.edu

**Introduction:**

Ferroelectric oxides combine switchable electric polarization with strong electromechanical coupling, making them indispensable in non-volatile memories, sensors, piezo-actuators and energy harvesters.[1] Their thin-film form is now ubiquitous in commercial ferroelectric random-access memories and radio-frequency components, while emerging uses span biomedical devices and photovoltaic conversion.[2] The push towards ultra-low-power electronics, non-volatile memories[3, 4], and beyond von Neumann logic architectures[5] has renewed interest in ferroelectrics that can sustain reliable switching[6] at nanometer dimensions and sub-femtojoule energies.[7-9]

Single-crystalline epitaxial films are especially attractive because coherent lattice order minimizes extrinsic losses and unlocks strain-engineered phases unavailable in polycrystalline counterparts.[10] Advances in oxide molecular-beam epitaxy, pulsed-laser deposition and metal–organic chemical-vapor deposition now permit unit-cell-level control of chemistry, defect density and epitaxial strain, enabling ferroelectric layers that tolerate biaxial strains of $\pm 3$ % without cracking.[10] Such misfit strains act as an internal "knob": compressive mismatch stabilizes $(001)$-oriented $c$ domains with out-of-plane polarization, whereas tensile mismatch favors $(100)/(010)$-oriented $a_1/a_2$ domains that polarize in-plane.[11] In $PbTiO_3$ and related $Pb(Zr,Ti)O_3$ films, tuning the strain field or the $Zr/Ti$ ratio can continuously rotate the polar axis, trigger $a \leftrightarrow c$ polymorph transitions and create mixed $a/c$ polydomain architectures that enhance piezoelectric response.[12] The resulting domain pattern, along with its variant fractions, wall densities, and ferroelastic twins, ultimately governs the macroscopic dielectric, piezoelectric, and switching behavior..[13-15]

Yet domain structures are far from homogeneous. Local variations in strain, electrostatics and defect chemistry cause wall mobility strongly site-dependent. Local structural inhomogeneities and topological defects can serve as preferential nucleation centers and wall-pinning sites. At electrode interfaces the built-in Schottky fields, misfit strain and broken symmetry lower the barrier to nucleation.[16] These hotspots can override even the highest applied fields at the film surface. Once a domain forms, its wall glides freely until it encounters rows of oxygen vacancies, heavy-cation substitutions, or impurity clusters.[17] Neutral walls lock onto local strain fields, while charged walls clamp onto ordered defect planes and halt mid-film.[18] The interplay between nucleation and wall pinning determines where and when domains form and stop, and hence define key device metrics such as coercive voltage, switching speed, piezoelectric response and fatigue lifetimes.[19]

Nanoscale scanning probe techniques such as piezoresponse-force microscopy (PFM) have been fundamental for visualizing these heterogeneities and for manipulating individual walls.[20-23] However, the density of the pinning centers is usually fairly low, limiting the potential of dense spectroscopic mapping methods. Similarly, human orchestrated experiment can select interesting regions, but can probe only a handful of locations, limiting statistical power.[24]

These considerations stimulated the development of automated, high-throughput variants of PFM.[25] Automated platforms like FerroBot use image-based triggers to apply bias pulses during PFM scans, enabling the controlled creation and probing of metastable domain states with minimal perturbation.[26, 27] Bayesian-optimization frameworks further extend these capabilities by guiding sparse sampling through Gaussian processes (GPs) that model both signal and noise, drastically reducing the number of measurements needed to map functional responses at microstructural elements.[25, 28-30] Hypothesis learning frames multiple competing physical models as Bayesian hypotheses with prior parameter distributions. It uses sequential PFM measurements to update model posteriors and choose the next bias-time condition to maximally discriminate between domain growth mechanisms.[31] In a $BaTiO_3$ switching study, this closed-loop strategy autonomously converged on domain-wall pinning kinetics as the dominant growth mechanism, albeit under the assumption that one of the predefined models fully captures the underlying physics.[31] Finally, combinatorial library approaches leverage continuous composition spreads and automated PFM read-outs to explore phase diagrams across multicomponent systems.[32, 33] While these advances demonstrate orders-of-magnitude speedups in sampling efficiency, the range of the explored phenomena has so far been limited by the single-line (FerroBot) or hysteresis loop detection.

Domain-wall dynamics in ferroelectric thin films are inherently position dependent, where each polar interface is locked into a distinct local microstructure. Correspondingly, application of bias pulses creates complex domain geometries that require direct visualization to explore. Conventional scanning probe methods lack the throughput to statistically sample these spatially sparse switching events, requiring manual intervention impractical. To overcome this limitation, we implemented a fully automated SPM pipeline in which Dual-AC Resonance Tracking PFM (DART-PFM) scans $10 \times 10$ $\mu m^2$ regions of epitaxial $PbTiO_3/KTaO_3$ to map both ferroelectric and ferroelastic domains, while a machine-learning–driven controller automatically executes and analyzes over a thousand bias-induced switching events. This automated workflow allows systematic exploration of the role of local pinning landscapes on domain wall dynamics.

**Materials and Experimental Objectives**

As a model system, epitaxial $PbTiO_3$ ($PTO$) thin films were deposited on (100) $KTaO_3$ single-crystal substrates by pulsed metal–organic chemical vapor deposition. Precursors of $Pb(C_{11}H_{19}O_2)_2$, $Ti(O-i-C_3H_7)_4$ and $O_2$ were pulsed in alternation, with film thickness tuned simply by varying deposition time. During growth the substrate was held at $600\ °C$ and $670\ Pa$; afterward it was cooled to room temperature at $10\ °C\ min^{-1}$. The coherent tensile misfit strain ($\sim 0.5\%$ in $50\ nm$ films, relaxing to $\approx 0\%$ by $1\ \mu m$ thickness) elevates the Curie temperature by $\sim 100\ °C$ and stabilizes in-plane polarization. To accommodate this tensile strain, $PTO$ forms a mixed polydomain pattern: narrow out-of-plane $\boldsymbol{c}$-domains interspersed within a predominantly in-plane $\boldsymbol{a}$-domain matrix, which itself splits into alternating $\boldsymbol{a_1}/\ \boldsymbol{a_2}$ ferroelastic twins separated by 90° walls. The spacing, volume fraction, and orientation of these $\boldsymbol{a_1}/\ \boldsymbol{a_2}$ twins and $\boldsymbol{c}$-lamellae vary strongly with local misfit, film thickness, and cooling conditions, producing a highly heterogeneous microstructural landscape upon which domain-wall pinning and switching dynamics are uniquely sensitive. The $\boldsymbol{c}$-domain regions in turn nucleate $\boldsymbol{a}$-domains, resulting in complex two-level domain pattern as illustrated in **Fig. 1a**.

To visualize the resulting ferroelectric and ferroelastic domain structure and quantify wall-switching behavior, the films were subsequently imaged by DART-PFM[34] over a micron-scale area of epitaxial $PTO$ with the resulting amplitude and phase contrast maps shown in **Fig. 1a** and **1b**, respectively. The microscope first acquired a high-resolution DART-PFM map of the entire $10 \times 10\ \mu m^2$ area (scan speed $25.04\ \mu m/s$, $2\ Hz$ line rate), then subdivided it into 100 contiguous $1 \times 1\ \mu m^2$ tiles. Within each tile, a pre-pulse PFM scan ($\sim 2.08\ min$) was immediately followed by on-the-fly Canny-edge filtering of the phase image to locate domain walls. For each experiment, five wall pixels per tile were randomly selected and subjected to five consecutive 10-second voltage pulses, all at either $10\ V$, $20\ V$, or $30\ V$. A matching post-pulse PFM map ($\sim 2.08\ min$) captured any local switching, and a brief stage move, and tip re-approach ($\approx 1\ min$) prepared the system for the next tile. In total, each $1 \times 1\ \mu m^2$ tile required $\sim 6.16\ min$, yielding 100 pre-post pulse map pairs and 500 targeted switching events per voltage magnitude. Finally, the complete ensemble of images and pulse metadata was assembled for subsequent clustering analysis of domain-wall switching behavior.

In a subsequent processing stage, $50\ nm \times 50\ nm$ subregions as shown in **Fig. 1** centered on each programmed pulse site were automatically extracted from the DART-PFM datasets. Within each subregion, the local ferroelectric polarization angle and the ferroelastic domain were quantified. The PFM image in **Fig. 1** demonstrates a complex hierarchy of polarization states, comprising six fundamental orientations ($a_1^+,\ a_1^-,\ a_2^+,\ a_2^-,\ c^-, c^+$) that combine to form twelve distinct domain patterns. These include four x-oriented walls ($a_1^+/c^-$ with $z^+$ twin walls, $a_1^+/c^+$ with $z^-$ twins, $a_1^-/c^-$ with $z^-$ twins, and $a_1^-/c^+$ with $z^+$ twins), four analogous y-oriented walls ($a_2{}^+/c^-$, $a_2{}^+/c^+$, $a_2{}^-/c^-$, $a_2{}^-/c^+$), and four in-plane configurations ($a_1{}^+/a_2{}^+$, $a_1{}^+/a_2{}^-$, $a_1{}^-/a_2{}^+$, $a_1{}^-/a_2{}^-$). While theoretical considerations suggest 132 possible domain boundary combinations, most of these walls are symmetry equivalent. Furthermore, experimentally we observe ferroelectric walls that have small polarization charge, or separate $a - c$ regions with opposite sense of twin wall tilts. Overall, experimental observations identify five dominant classes organized by their crystallographic alignment and polarization transitions. Crystallographic-aligned walls (Class I) parallel to principal planes, exemplified by $a_2{}^+/c^+ \parallel a_2{}^-/c^-$ configurations in **Fig. 1c**, demonstrate co-linear ferroelectric-ferroelastic alignment, while orthogonal walls (Class II) perpendicular to principal planes, shown in **Fig. 1e** as $a_2{}^+/c^+ \parallel a_2{}^-/c^-$ boundaries, exhibit pure 90° ferroelastic rotation. Angled walls (Class III) inclined at $\boldsymbol{\theta}$ angles relative to crystallographic axes are visible in **Fig. 1d** as $a_2{}^+/c^+ \parallel a_2{}^-/c^-$ boundaries, where domain walls follow local microstructure-dependent orientations. The $a_1c\ /\ a_2c$ boundaries (Class IV) transitioning between $x-$ and $y-$ oriented domains appear in **Fig. 1f** as $a_1{}^-/c^+ \parallel a_2{}^-/c^-$ interfaces, whereas $a_1c\ /\ a_{12}c$ boundaries (Class V) separating $a/c$ and in-plane domains are illustrated in **Fig. 1g** by $a_1{}^-/a_2{}^- \parallel a_2{}^-/c^-$ configurations. These experimentally observed domain configurations are systematically classified into five distinct groups (Classes **I-V**), which form the basis for our subsequent analysis. The complete set of ferroelectric/ferroelastic orientation relationships, along with their defining angular characteristics, are visually presented in **Fig. 1c-g** and quantitatively summarized in **Table 1**.

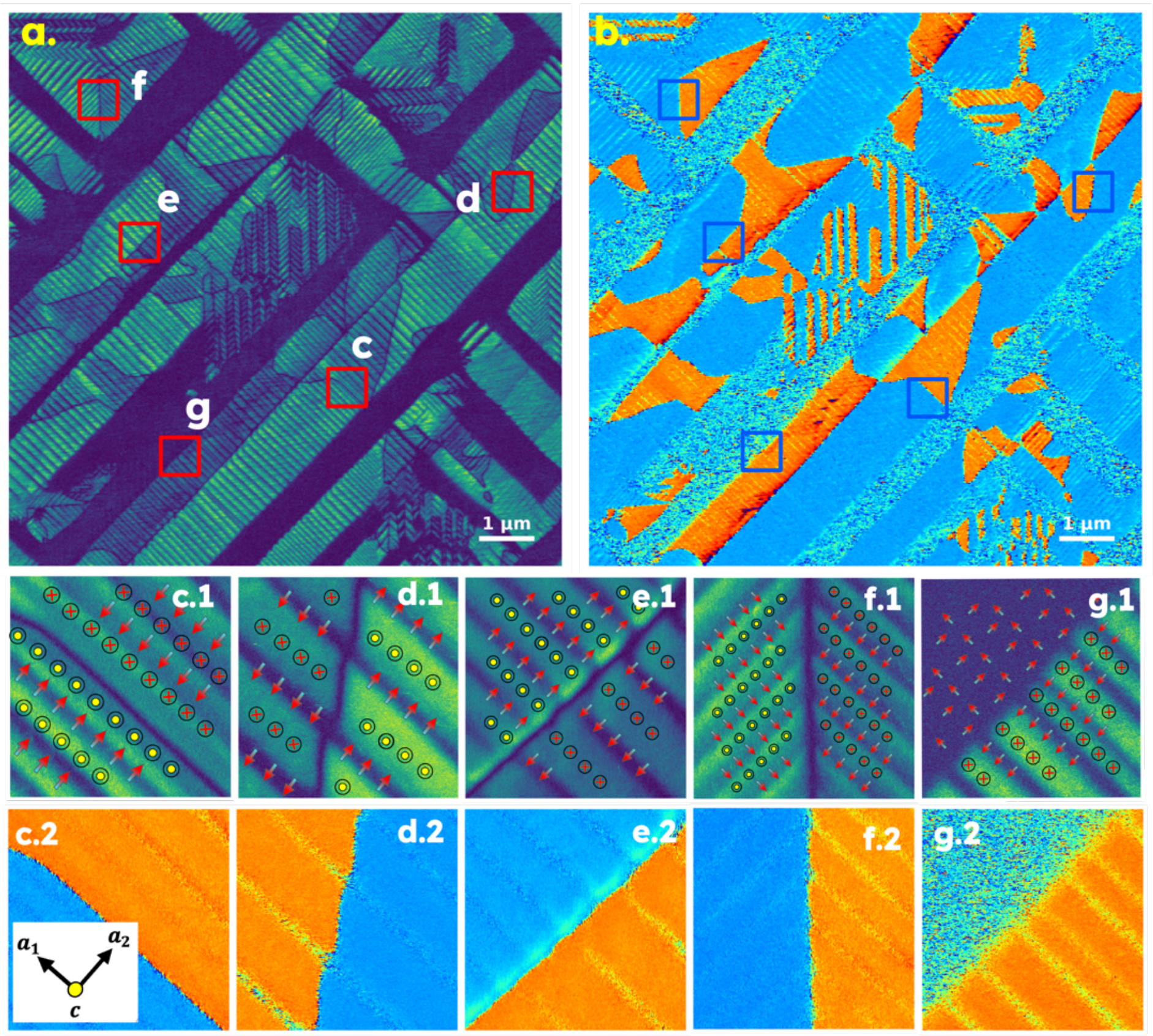

**Figure 1**: **a.** Amplitude channel acquired over a $10\mu m * 10\mu m$ region, showing the structural contrast associated with ferroelectric and ferroelastic domain walls, **b.** Corresponding phase channel, selectively highlighting ferroelectric domain walls. Subplots (**c.1** – **g.1**) on amplitude channel and (**c.2** – **g.2**) on phase channel present representative local regions exhibiting distinct combinations of ferroelectric and ferroelastic domain wall configurations in bisects of $1\mu m * 1\mu m$, illustrating the diversity of domain wall interactions captured in the scanned area.

**Table 1:** Clustering definition of domain coupled structure in PTO/KTO materials

| Class | Fig. panels | Characteristics |
|---|---|---|
| $\boldsymbol{Class\ I}$ | c.1, c.2 | $\boldsymbol{a_2^+/c^+ \parallel a_2^-/c^-}$ |
| $\boldsymbol{Class\ II}$ | d.1, d.2 | $\boldsymbol{a_2^+/c^+ \parallel a_2^-/c^-, angled\ at\ \theta°}$ |
| $\boldsymbol{Class\ III}$ | e.1, e.2 | $\boldsymbol{a_2^+/c^+ \parallel a_2^-/c^-, angled\ at\ 90°}$ |
| $\boldsymbol{Class\ IV}$ | f.1, f.2 | $\boldsymbol{a_1^-/c^+ \parallel a_2^-/c^-}$ |
| $\boldsymbol{Class\ V}$ | g.1, g.2 | $\boldsymbol{a_1^-/a_2^- \parallel a_2^-/c^-}$ |

**Domain wall Segmentation and Orientation Mapping**

Extracting precise ferroelastic wall orientations from PFM data requires building a multi-stage image-processing pipeline designed to suppress noise, enhance true ridge features, and unambiguously separate ferroelastic from ferroelectric boundaries. In the unprocessed DART-PFM amplitude channel **Fig. 2a**, ferroelectric switching walls and ferroelastic tilt boundaries are superimposed on uneven intensity fields arising from tip–sample coupling and local surface topography.[35] Direct gradient-based orientation fitting under these conditions yields noisy, unreliable results. To enhance the true wall ridges, the amplitude map is processed by a Sato filter[36], a vessel-enhancement operator originally developed for biological image segmentation where elongated, ridge-like structures must be accentuated. The Sato filter works by analyzing the second-order derivatives of the image at multiple scales. At each scale $\sigma$, one computes the Hessian matrix:

$$H_\sigma(x) = \begin{pmatrix} I_{xx}(x,\sigma) & I_{xy}(x,\sigma) \\ I_{yx}(x,\sigma) & I_{yy}(x,\sigma) \end{pmatrix} \quad (1)$$

where $I_{ij}(x,\sigma)$ are the Gaussian-smoothed second derivatives of the amplitude map. Its eigenvalues $\lambda_1, \lambda_2$ (with $|\lambda_1| \leq |\lambda_2|$) describe local curvature. A ridge-enhancement response at scale $\sigma$ is then defined as:

$$R = \frac{\lambda_1}{\lambda_2}, \quad S = \sqrt{\lambda_1^2 + \lambda_2^2} \quad (2)$$

$$V_\sigma(x) = \begin{cases} 0, & \lambda_2 > 0 \\ \exp\left(-\frac{R^2}{2\alpha^2}\right)[1 - \exp(-\frac{S^2}{2\beta^2}), & otherwise \end{cases} \quad (3)$$

where $\alpha$ and $\beta$ control sensitivity to anisotropy and contrast. The final enhanced map is the maximum response across a set of scales:

$$V(x) = \max_\sigma V_\sigma(x) \quad (4)$$

Where it highlights elongated, line-like features while suppressing isotropic noise, producing a high-contrast "ridge" image as presented in **Fig. 2b**. A structure-tensor computation then converts these ridges into a continuous orientation field, whose most reliable segments are isolated by thresholding the response $V_\sigma$ at its $90th$ percentile **Fig. 2c**.

In parallel, the PFM phase channel is processed with Canny edge detection[37] to pinpoint only the 180° ferroelectric boundaries **Fig. 2d–e**. By masking out every phase-detected pixel from the ridge-derived orientation field—i.e., retaining orientation vectors only where no ferroelectric edge exists—the workflow generates a ferroelastic orientation map represented in **Fig. 2f**.

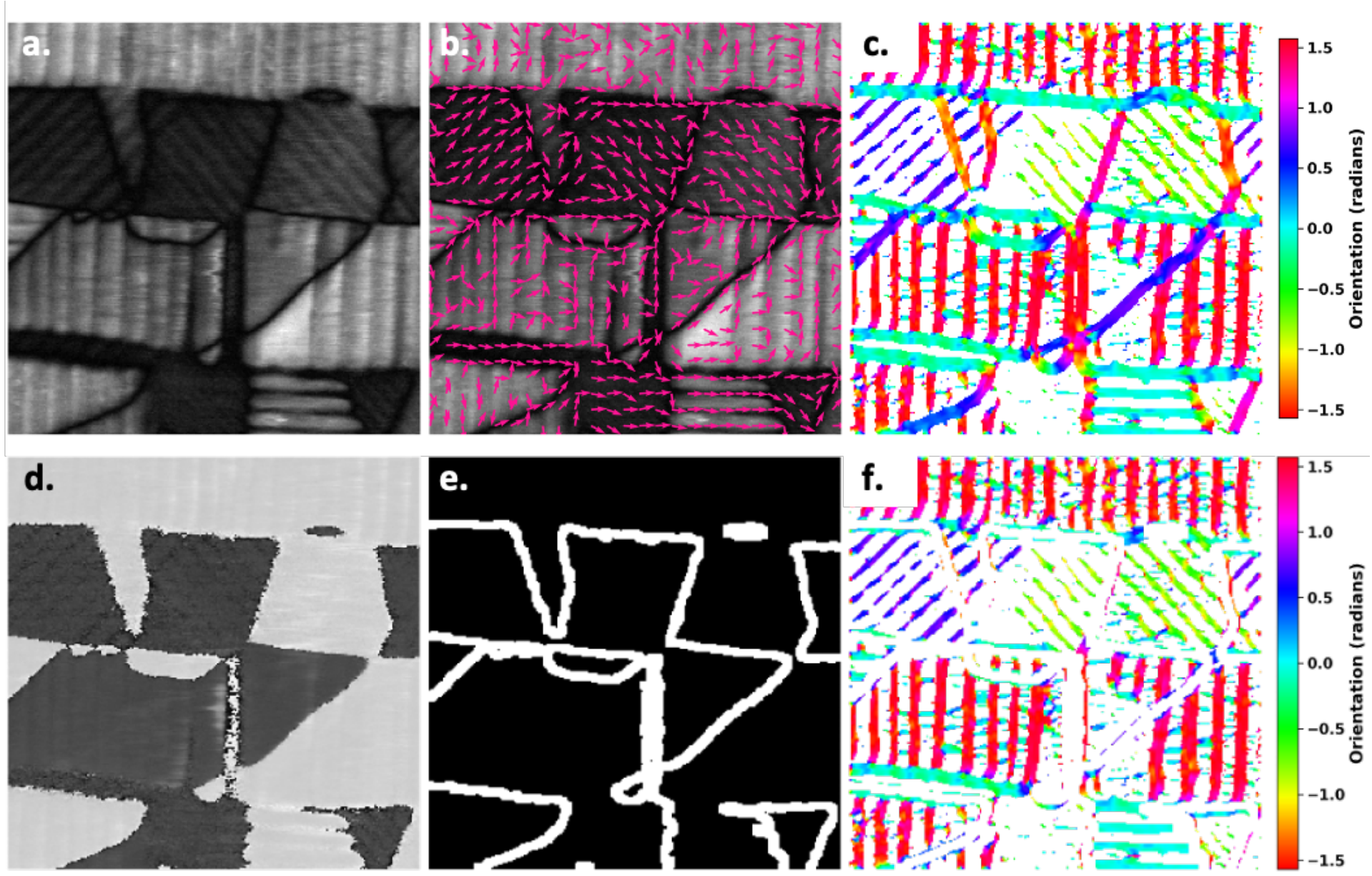

**Figure 2**: Ordered analysis steps for domain structure characterization, **a.** Amplitude channel revealing ferroelastic and ferroelectric domain structures, **b.** Local orientation field derived via structure tensor analysis of the Sato-filtered amplitude channel, **c.** Orientation map highlighting all domain walls orientations after percentile-based thresholding, **d.** Phase channel offering complementary contrast for ferroelectric domain boundaries, **e.** Ferroelectric domain wall mask obtained via Canny edge detection on the phase channel, **f.** Filtered orientation map isolating ferroelastic domains by masking ferroelectric walls. Patch sizes are $3\mu m$.

To explore the relationship between ferroelectric wall dynamics and ferroelastic domain orientation, we proceed as following. At each pulse site $P_k$ a $50\,nm \times 50\,nm$ square neighborhood $N_S(P_k)$ was defined, and the ferroelastic orientation values $\theta_{fe}(x, y)$ within that window were sampled. These local angles were aggregated into an orientation histogram $h_k(\varphi)$ spanning $[0, \pi)$. A Gaussian mixture model[38] was then fit to $h_k(\varphi)$, with the number of components $J_k$ selected by minimizing the Bayesian Information Criterion. Each component $J$ is described by a weight $\omega_{k,j}$, mean orientation $\mu_{k,j}$, and variance $\sigma_{k,j}^2$. The dominant orientation $\varphi_k^*$ in each patch was taken to be the component mean $\mu_{k,j^*}$ associated with the largest weight $\omega_{k,j}$. Finally, each $\varphi_k^*$ was assigned to the nearest archetypal wall angle (0°, $\theta$°, or 90°), as defined by the five domain-wall clusters (**Table 1**, **Fig. 1c–g**), yielding a fully labeled dataset linking local wall geometry to switching behavior. The entire pipeline is distilled in **Algorithm 1**.

**Algorithm 1:** Patch-Based Orientation Analysis for Ferroelastic Walls

**Input:**

1. **Amplitude Channel :** $\boldsymbol{I_1} : \boldsymbol{\Omega} \to \mathbb{R}$
2. Phase channel: $\boldsymbol{I_2} : \boldsymbol{\Omega} \to [\boldsymbol{0}, \boldsymbol{2\pi})$
3. Pulse Locations: $\{\boldsymbol{p_k}\}_{\boldsymbol{k=1}}^{\boldsymbol{K}} \subset \boldsymbol{\Omega}$

**Output:** Set of patch orientation labels: $\{\boldsymbol{\varphi_k^*}\}_{\boldsymbol{k=1}}^{\boldsymbol{K}}$

**Steps:**

1. Apply Sato filtering to $\boldsymbol{I_1}$ to obtain all orientations within channel: $\boldsymbol{I_1^{Sato}} : \boldsymbol{\Omega} \to \mathbb{R}$
2. Compute the local orientation map from $\boldsymbol{I_1^{Sato}}$: $\boldsymbol{\theta(x,y)} : \boldsymbol{\Omega} \to [\boldsymbol{0}, \boldsymbol{\pi})$
3. Apply Canny edge detection to phase channel $\boldsymbol{I_2}$ to obtain only ferroelectric walls: $\boldsymbol{M_{FE}} : \boldsymbol{\Omega} \to \{\boldsymbol{0}, \boldsymbol{1}\}$

Where:

$\boldsymbol{M_{FE}(x,y) = 1}$, If a ferroelectric wall is detected at (x, y)

$\boldsymbol{M_{FE}(x,y) = 0}$, Otherwise

4. Construct the masked orientation map to keep only ferroelastic walls by zeroing out the ferroelectric locations:

$$\boldsymbol{\theta_{fe}(x,y)} = \begin{cases} \boldsymbol{\theta(x,y)}, & \boldsymbol{if\ M_{FE}(x,y) = 0} \\ \boldsymbol{Undefined}, & \boldsymbol{if\ M_{FE}(x,y) = 1} \end{cases}$$

5. For each pulse location: $\boldsymbol{P_k} \in \boldsymbol{\Omega}$

i. Extract local orientation patch:

$$\boldsymbol{P_k} = \{\boldsymbol{\theta_{fe}(x,y)} : \boldsymbol{(x,y)} \in \boldsymbol{N_S(P_k)}\}$$

Where:

$\boldsymbol{N_S(P_k)}$ is the square neighborhood of side $\boldsymbol{s}$ centered at $\boldsymbol{P_k}$

i. Compute the orientation histogram $\boldsymbol{h_k(\varphi)}$ over defined orientations in $\boldsymbol{P_k}$

i. Fit Gaussian Mixture Model (GMM) to $\boldsymbol{h_k(\varphi)}$ as

$$\boldsymbol{h_k(\varphi)} = \sum_{\boldsymbol{j=1}}^{\boldsymbol{J_k}} \boldsymbol{\omega_{k,j} N(\varphi;\ \mu_{k,j}, \sigma_{k,j}^2)}$$

/. Identify the dominant orientations: $\boldsymbol{\varphi_k^*} = \boldsymbol{\mu_{k,j^*}}$, where $\boldsymbol{j^*} = \mathbf{argmax}_{\boldsymbol{j}}\ \boldsymbol{\omega_{k,j}}$

6. Return the set of orientation labels $\{\boldsymbol{\varphi_k^*}\}_{\boldsymbol{k=1}}^{\boldsymbol{K}}$

**Microstructure-Resolved Switching Dynamics**

By leveraging a large dataset of over 1,500 paired pre- and post-pulse DART-PFM amplitude and phase images (comprising approximately 500 for each voltage of $10\ V$, $20\ V$, and $30\ V$), each centered on a preexisting 180° ferroelectric wall and grouped into five distinct wall-orientation clusters, we achieve the statistical power needed. This enables us to disentangle true structure-dependent switching from local variability. In every experiment, voltage pulses are applied precisely at the native wall center of the $50\ nm \times 50\ nm$ patch as presented in Fig 3a ($\boldsymbol{A_0}$), precluding artifacts from newly nucleated boundaries and isolating the intrinsic response. To quantitatively compare the switching response of each cluster, the primary evaluation metric was defined as the pixel-wise absolute difference between pre- and post-pulse PFM images, both amplitude and phase channels. **Fig. 3a-c** show the raw amplitude ($\boldsymbol{A_0}$) and phase ($\boldsymbol{P_0}$) maps acquired just before pulsing, alongside the corresponding post-pulse maps ($\boldsymbol{A_1}, \boldsymbol{P_1}$). Subtracting these yields the high-contrast difference maps ($\boldsymbol{A_d}$), which spatially resolve nanoscale wall displacement, polarization reorientation, and domain expansion. This rigorously derived difference field then underpin the mobility analysis presented in **Fig. 4** and **5**, revealing how specific ferroelastic architectures modulate both the critical switching bias and the creep-to-unpin dynamics in a scientifically precise, architecture-resolved manner.

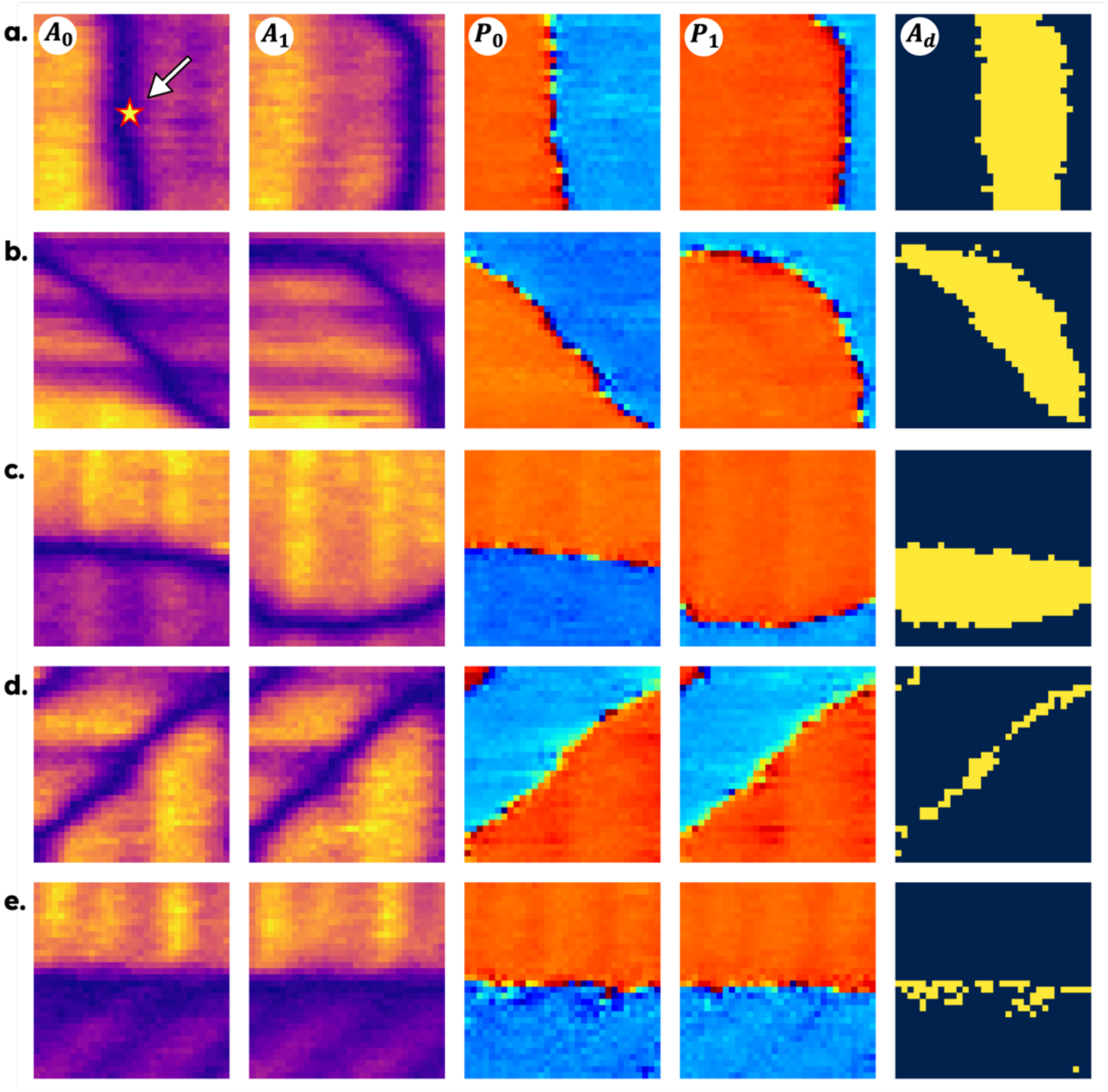

**Figure 3**: Class-dependent domain-wall response to an applied voltage pulse, each row corresponds to one of the five wall-space classes (from up to bottom: $classes\ \ I\ to\ V$). Within each row, the five panels show: $A_0$: amplitude before the pulse (star marks the pulse application site at the wall center), $A_1$: amplitude after the pulse, $P_0$: phase before the pulse, $P_1$: phase after the pulse, $A_d$: thresholded, absolute phase difference, highlighting the domain-wall displacement induced by the pulse.

Uniform ferroelastic structure, where the ferroelectric switching wall bisects two ferroelastic domains that both lie along the same crystallographic direction ($\boldsymbol{Class\ I, II, and\ III}$), exhibit a classic single-step activation. As shown in **Fig. 4**, under a $10\ V, 10\ s$ pulse they remain essentially static ($< 15\ \%$ pixel displacement), but once the applied field exceeds the local coercive threshold at $20\ V$, wall segments jump by $\sim\ 20–25\ \%$, rising further to $\sim 26–28\ \%$ at $30\ V$. By contrast, heterogeneous ferroelastic boundaries ($\boldsymbol{Class\ IV, and\ V}$), in which the ferroelectric trace separates two ferroelastic domains of different orientations, remain largely static at both $10\ V$ and $20\ V$, indicating a strong pinning potential. This intermediate plateau reflects an elastic/electrostatic pinning imposed by the misaligned ferroelastic lattice on either side of the wall. Only above $\sim\ 25\ V$ does this secondary barrier yield, unleashing a rapid displacement to $> 20\ \%$ at $30\ V$ and aligning with the uniform-boundary response. These structure-specific "mobility fingerprints" demonstrate that the relative orientation of adjacent ferroelastic domains controls the field-dependence of wall motion, either a single coherent threshold or a two-step creep-and-unpin sequence, offering quantitative design rules for targeted domain manipulation in ferroelectric devices.

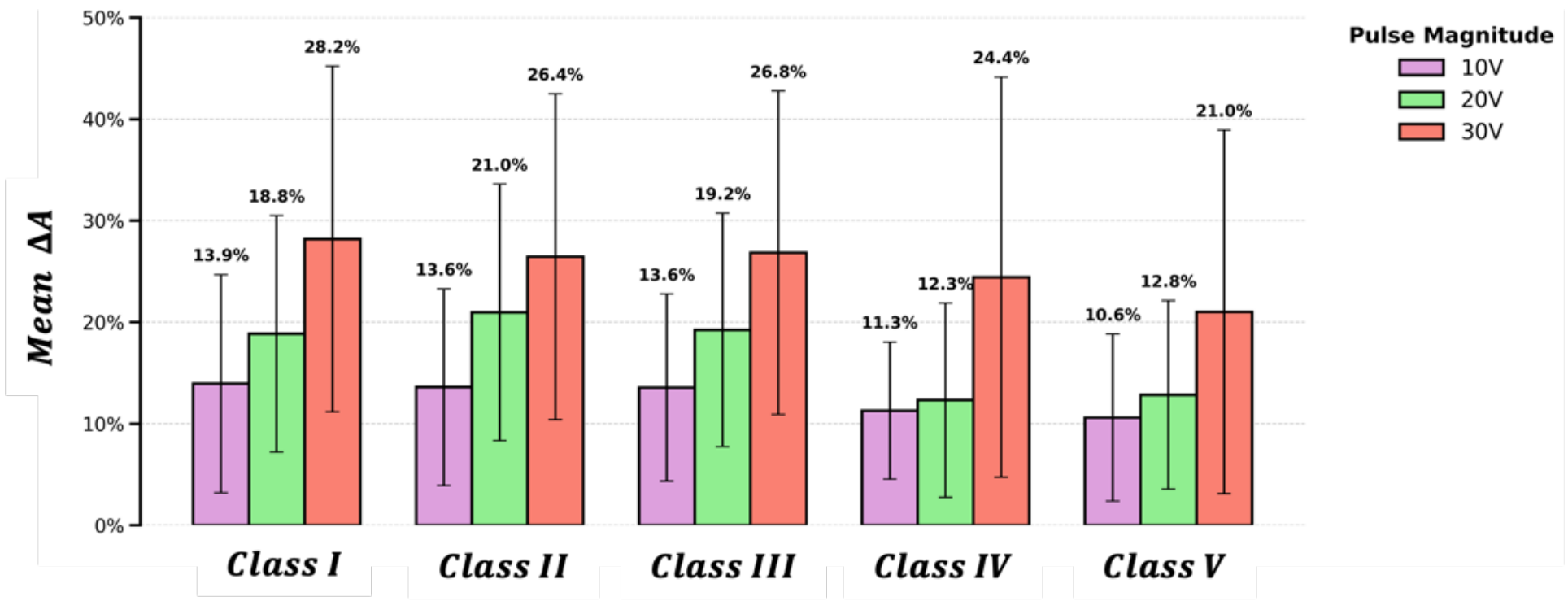

**Figure 4.** Dependence of ferroelastic wall displacement on pulse magnitude for each orientation cluster. Mean pixel-wise absolute displacement (expressed as a percentage of the 50 nm × 50 nm patch area) for each of the five wall-orientation clusters after $10\ s$ pulses at $10\ V$, $20\ V$, and $30\ V$. Error bars denote one standard deviation across all sampled patches.

At high field $30\ V$, the angular trend shown in **Fig. 5** further supports this conclusion: domain wall displacements reach local maxima near ~50° and ~140°, corresponding to intersection geometries that facilitate relaxation through minimized electrostatic and elastic mismatch. In contrast, sharp minima at ~40°, ~80°, and ~155° reflect orientations where the ferroelectric wall is pinned, likely due to polarization frustration or strain incompatibility across the interface. These energy barriers suppress wall motion even at elevated bias, underscoring that domain switching at $30\,V$ is still highly sensitive to local wall geometry. Together, these results affirm that angular alignment encodes structure-specific energy landscapes that shape switching behavior under applied field.

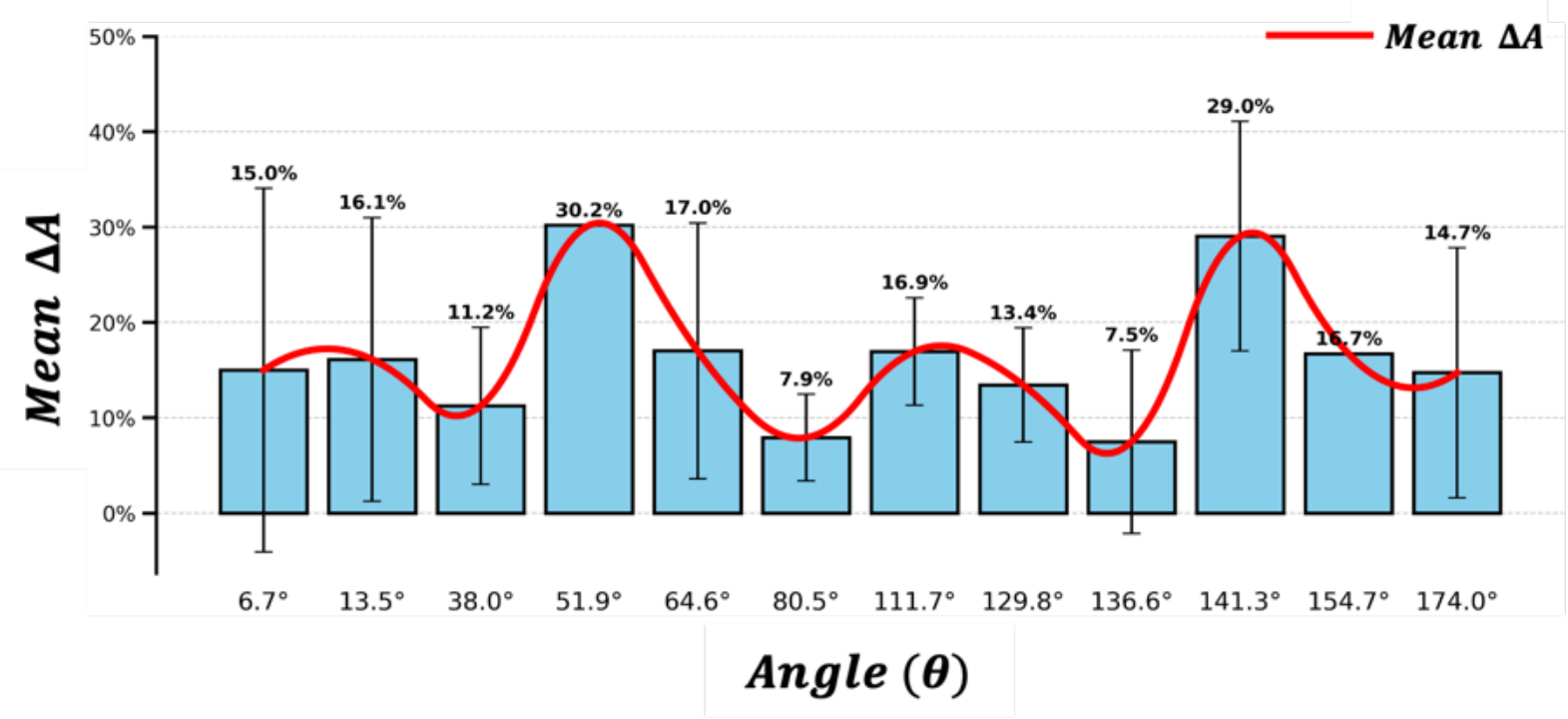

**Figure 5:** Domain-wall mobility as a function of angular offset between ferroelectric trace and ferroelastic tilt.

To summarize, we have developed an automated SPM workflow that combines large-area DART-PFM mapping with on-the-fly domain-wall detection and ML-driven patch extraction to explore ferroelectric wall dynamics at scale. This approach to massively accelerate exploration of the microstructure-dependent switching dynamics, allowing to explore 1500 switching events in 30 hours. The equivalent human-orchestrated experiment would have taken 150 events, thus suggesting the acceleration of 10-fold via automated SPM while also eliminating human bias and error. This high-throughput data reveals two distinct mobility fingerprints: uniform ferroelastic walls switch coherently above a single threshold, while heterogeneous walls exhibit a two-step creep-and-unpin behavior.

Here the focus of the experiment was on epitaxial $PbTiO_3/KTaO_3$ films with a finite set of five wall orientations. Extending this approach to materials with more complex, polycrystalline microstructures will be enabled by integrating EBSD mapping of Euler angles with automated domain-identification workflows. Such multimodal datasets will allow us to chart domain-wall pinning at grain boundaries, probe interactions between different wall classes, and capture domain-wall reactions under bias.[10, 39] Quantitative PFM methods on Vero AFMs[40] that employ patented Quadrature Phase Differential Interferometry (QPDI)[31] to measure true tip displacement ensure amplitude and phase shifts are calibrated with unparalleled precision, and repeatability. Looking ahead, we envision a roadmap as shown in **Fig. 6**, that illustrates how these combined techniques can systematically explore wall behavior across grain boundaries, heterointerfaces, and complex twin networks.

More broadly, closing the characterization loop in functional materials demands autonomous, data-rich experiments that fuse complementary probes (PFM, EBSD, and TEM) and in situ stimuli. This self-driving labs paradigm makes it possible to learn dynamic mechanisms in solids on local length and time scales.[26, 41, 42] By embedding real-time analysis and decision-making into the microscope workflow, we can accelerate the discovery of structure–property rules and feed them back into synthesis and device design. Ultimately, this integrated vision of multimodal imaging, quantitative PFM, and machine-learning-guided experimentation will transform our ability to engineer reliable, low-energy ferroelectric memories, actuators, and sensors.

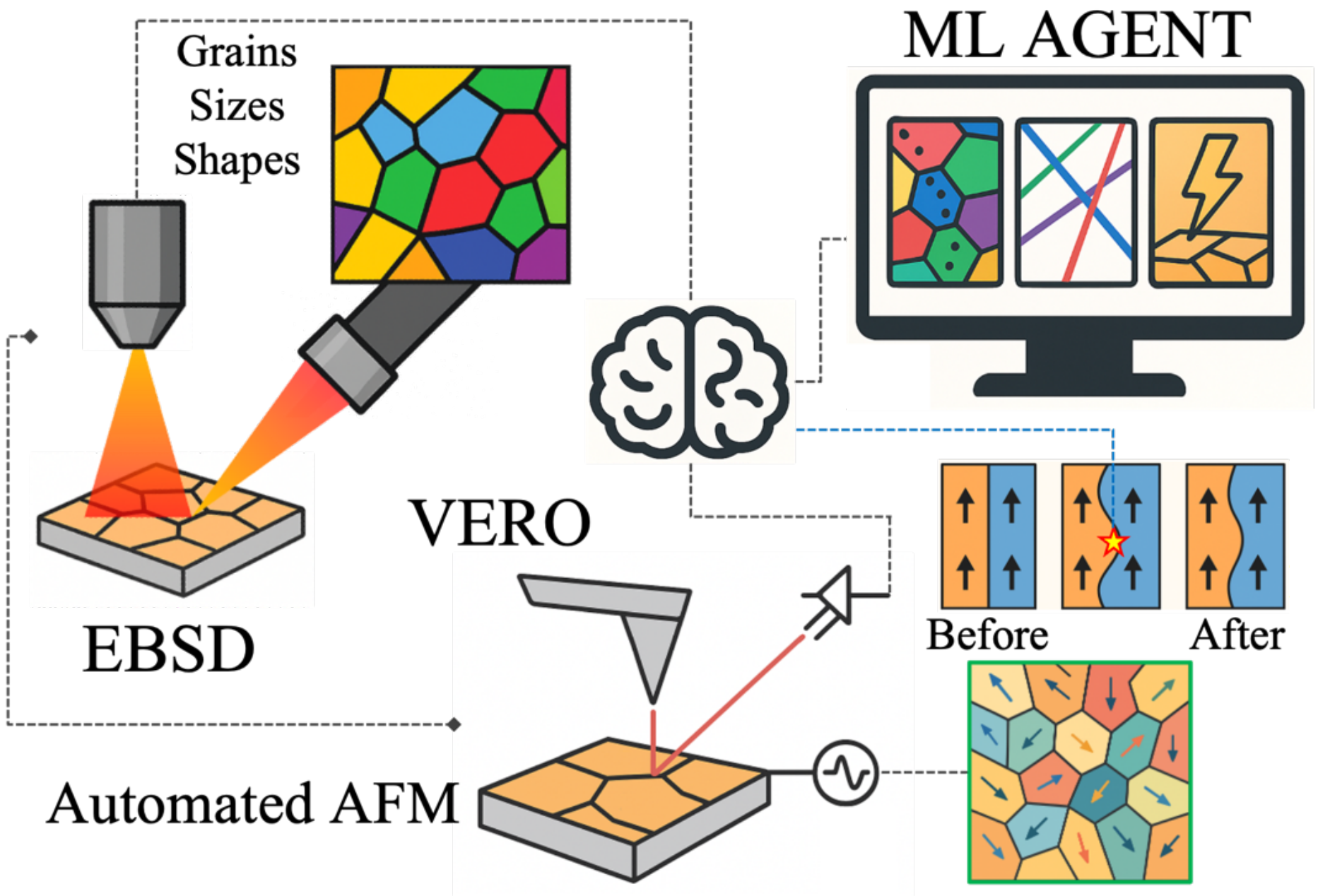

**Figure 6**: Integrated AI-driven workflow for microstructure-resolved ferroelectric switching

**ACKNOWLEDGMENTS**:

This work (workflow development) was supported (K.B., S.V.K.) by the U.S. Department of Energy, Office of Science, Office of Basic Energy Sciences Energy Frontier Research Centers program under Award Number DE-SC0021118." H. F. was partially supported by the Japan Science and Technology Agency (JST) as part of Adopting Sustainable Partnerships for Innovative Research Ecosystem (ASPIRE)(JPMJAP2312), MEXT Initiative to Establish Next-generation

Novel Integrated Circuits Centers (X-NICS) (JPJ011438), and MEXT Program: Data Creation and Utilization Type, Material Research and Development Project (JPMXP1122683430).

## AUTHOR DECLARATIONS

Conflict of Interest: The authors have no conflicts to disclose.

**Author Contributions:**

Kamyar Barakati: Conceptualization (equal), Data curation (lead), Formal analysis (equal), Writing – original draft (equal), Methodology (equal); Sergei V. Kalinin: Conceptualization (equal), Formal analysis (equal), Funding acquisition (equal), Writing – review & editing (equal), Supervision (equal); Yu Liu: Software (equal), Investigation (equal): Hiroshi Funakubo: Methodology (equal), Resources (equal), Investigation (equal).

## DATA AVAILABILITY:

The data that support the findings of this study are available from the corresponding authors upon request.